\documentclass[twocolumn, iop, apj, twocolappendix, numberedappendix, appendixfloats]{emulateapj}
\usepackage{CJKutf8}
\usepackage{threeparttable}
\usepackage{multirow}
\usepackage{times}
\usepackage{enumitem}
\usepackage{url} 
\usepackage{color}
\usepackage{amsmath,bm}
\usepackage{subfigure}
\usepackage{natbib}
\usepackage{graphicx}
\usepackage{graphics}
\usepackage{hyperref}                             
\usepackage{amsmath}
\usepackage{bm}
\usepackage{subfigure}
\usepackage{array}
\usepackage{booktabs}
\usepackage{lineno}

\shorttitle{Radial Metallicity Gradients in the Thin and Thick Disks}

\shortauthors{Sun et al.}
\submitted{Submitted to ApJL;\,\,\,\,Accepted July 6, 2026}

\begin{document}
%\linenumbers

\title{Divergent Evolution of Radial Metallicity Gradients in the Thin and Thick Disks of the Milky Way}

\author{Weixiang Sun\textsuperscript{1,7}}

\author{Gaohuan Long\textsuperscript{2,3}}

\author{Hui Li\textsuperscript{1,7}}

\author{Han Shen\textsuperscript{4,5}}

\author{Shu Wang\textsuperscript{3}}

\author{Xiaodian Chen\textsuperscript{3}}

\author{Biwei Jiang\textsuperscript{2}}

\author{Xiaowei Liu\textsuperscript{6,7}}

\author{Di Li\textsuperscript{1}}

\altaffiltext{1}{Department of Astronomy, Tsinghua University, Beijing 100084, People’s Republic of China; {\it sunweixiang@tsinghua.edu.cn {\rm (WXS)}}; 
{\it hliastro@tsinghua.edu.cn{\rm (HL)}}}

\altaffiltext{2}{School of Physics and Astronomy, Beijing Normal University, Beijing 100875, People’s Republic of China}

\altaffiltext{3}{CAS Key Laboratory of Optical Astronomy, National Astronomical Observatories, Chinese Academy of Sciences, Beijing 100101, People's Republic of China}

\altaffiltext{4}{School of Physics, University of New South Wales, Kensington 2032, Australia}

\altaffiltext{5}{ARC Centre of Excellence for All Sky Astrophysics in 3 Dimensions (ASTRO 3D), Australia}

\altaffiltext{6}{South-Western Institute for Astronomy Research, Yunnan University, Kunming 650500, People's Republic of China; {\it x.liu@ynu.edu.cn {\rm (XWL)}}}

\altaffiltext{7}{Corresponding authors}

\begin{abstract}

Using 200,388 red clump stars from LAMOST and APOGEE, we investigate the radial metallicity gradients of the Galactic disk as a function of vertical height and stellar age.
The thin disk displays a pronounced negative radial metallicity gradient near the Galactic mid-plane that progressively flattens with increasing $|Z|$, following $\Delta \mathrm{[Fe/H]}/\Delta R$ = $-$0.0784\,$+$\,0.0776\,(1\,$-$\,exp\,($-$\,$|Z|$\,/\,1.42)).
The thin disk also exhibits a clear age dependence in radial metallicity gradients, evolving smoothly from a strong gradient regime for young stars to a weak gradient regime for old stars, following $\Delta \mathrm{[Fe/H]}/\Delta R$ = $-$0.0438 $+$ 0.0233\,tanh\,(($\tau$\,$-$\,11.29)\,/\,4.21).
The thick disk shows weakly positive radial metallicity gradients that remain statistically invariant with respect to both vertical height and stellar age, following respectively, $\Delta \mathrm{[Fe/H]}/\Delta R$  = 0.0038\,$+$\,0.0009\,$|Z|$ and $\Delta \mathrm{[Fe/H]}/\Delta R$  = 0.0146\,$-$\,0.0007\,$\tau$.
These results indicate that the thin disk retains radial metallicity gradients shaped by relatively ordered inside-out growth and long-term secular evolution processes.
The thick disk exhibits spatially and temporally homogeneous radial metallicity gradients, which are consistent with a formation environment characterized by mergers of gas-rich systems and/or the turbulent ISM.

\end{abstract}


\section{Introduction}

Understanding the formation and evolution of the Galactic disk is one of the central problems in Galactic astronomy and galaxy formation theory \citep[e.g.,][]{Matteucci2012, Bland-Hawthorn2016}. 
Radial metallicity gradients provide important observational constraints on these processes because they preserve rich information about numerous complex physical processes in the Milky Way, including: star formation and chemical evolution \citep[e.g.,][]{Matteucci1989, Schonrich2017, Grisoni2018}, stellar migration \citep[e.g.,][]{Sellwood2002, Lee2011, Sun2024a}, the infall and outflow of gas \citep[e.g.,][]{Larson1976, Pezzulli2016, Andrews2017}, the accretions of dwarf galaxies \citep[e.g.,][]{Quinn1993}, as well as the non-axisymmetric perturbations induced by the central bar or spiral arms \citep[e.g.,][]{Andrievsky2004, Scarano2013}.

The radial metallicity gradients have been widely characterized by various tracers, including Cepheids \citep[e.g.,][]{Andrievsky2002, Luck2006}, open clusters \citep{Chen2003, Magrini2009}, planetary nebulae \citep{Costa2004, Henry2010}, OB stars \citep{Daflon2004}, H$_\mathrm{II}$ regions \citep{Balser2011}, FGK dwarfs \citep{Katz2011, Boeche2013}, giant stars \citep{Hayden2014, Yan2019}, red clump star \citep{Huang2015}, as well as main-sequence turn-off and sub-giant (MSTO-SG) stars \citep{Wang2019}.
These results indicate that the disk stars have negative radial metallicity gradients, and exhibit significant spatial dependence.
These observed gradients are in good agreement with the prediction of the classical ``inside-out” star formation scenario, which suggests that the establishment of negative radial metallicity gradients as a consequence of the progressive growth of the Galactic disk \citep[e.g.,][]{Chiappini2009, Schonrich2017, Frankel2019}.

Several studies determined the radial [Fe/H] gradients of mono-age stellar populations, and found that the radial gradients show strong variations among different mono-age stellar populations \citep[e.g.,][]{Xiang2015, Wang2019}.
The radial gradients are generally flat for the oldest populations, and become steeper with decreasing age, and then flatten again as age decreases.
These results reveal that subsequent secular disk evolution, including radial migration, dynamical heating, and disk mixing, can redistribute stars and partially erase primordial metallicity gradients \citep[e.g.,][]{Lee2011, Minchev2013, Mackereth2019}. 
Disentangling the relative contributions of these processes remains a major challenge for understanding the chemo-dynamical evolution of the Milky Way disk.

\begin{figure*}[t]
\begin{center}
\includegraphics[width=17.8cm]{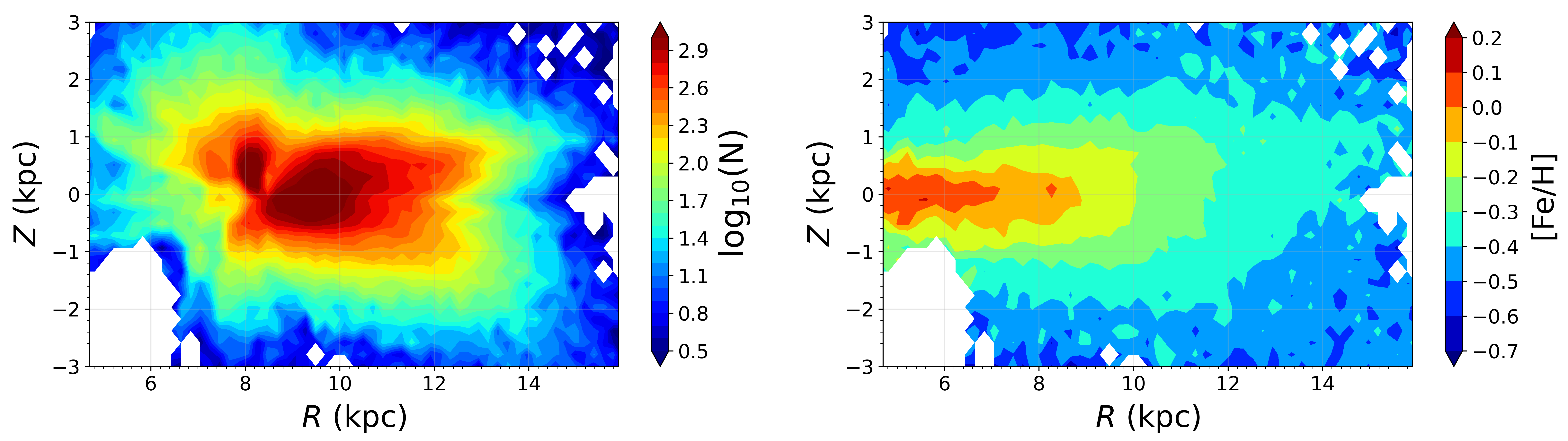}

\caption{{\bf Left panel:} Spatial distribution in the $R$ - $Z$ plane, of the whole sample stars, with color-coded by the stellar number density.
There is a minimum of 5 stars per bin spaced 0.2\,kpc in both axes.
{\bf Right panel:} Similar to the left-hand panel but color-coded by stellar [Fe/H].
}
\end{center}
\end{figure*}
%%\label{fig1}

Large spectroscopic surveys in the new era have enabled increasingly detailed measurements of the metallicity structure of the Galactic disk \citep[e.g.,][]{Boeche2014, Hawkins2023, Imig2023}. 
Previous studies have shown that radial metallicity gradients vary systematically with Galactic position and stellar population \citep[e.g.,][]{Yan2019, Han2020, Sun2024a}. 
In particular, the thin and thick disks exhibit markedly different chemical and kinematic properties, suggesting distinct evolutionary histories \citep[e.g.,][]{Lee2011, Bensby2014, Sun2024b}.
The thin disk generally exhibits ordered chemical structures consistent with long-term disk evolution, whereas the thick disk likely formed during an earlier and dynamically hotter phase of Galactic evolution \citep[e.g.,][]{Brook2012, Wisnioski2015, Sun2024a, Sun2025a}.

However, a detailed analysis of the coupled spatial and temporal behavior of radial metallicity gradients for chemically distinct stellar populations is still not yet well measured over a larger disk volume since the samples used in previous studies lack accurate measurement of the stellar age, chemistry and distance \citep[e.g.,][]{Lee2011, Yan2019, Sun2024a}, which would enable a credible assessment of the disk assembly histories.
In particular, it is still unclear whether the thick disk preserves measurable radial metallicity gradients across different vertical heights and stellar ages, or whether efficient early mixing processes largely wipe out these gradients.

At present, the larger sample of red clump (RC) stars \citep[e.g.,][]{Bovy2014, Huang2020, Wang2021} selected from LAMOST \citep[e.g.,][]{Cui2012, Deng2012} and APOGEE \citep{Majewski2017} surveys, presents an excellent opportunity to study this field. 
Based on this sample, it is possible to conduct an exhaustive study of the degree to which radial metallicity gradients vary with both vertical height and stellar age, therefore providing a direct observational probe of the persistence or erasure of chemical structures in the Galactic disk, thereby constraining the disk evolution histories.

This paper is structured as follows.
We describe the data used in Section 2, and present results and discussion in Section 3.
Finally, our main conclusions are summarized in Section 4.

\section{Data}

In this paper, we mainly used a sample with 200,388 RC stars selected from the APOGEE \citep{Majewski2017} and LAMOST \citep[e.g.,][]{Cui2012, Deng2012} spectroscopic surveys following the RC selection methods of \citep{Bovy2014} and \citep{Huang2020}, selecting the parameters with a large signal-to-noise ratio (SNR) for the common sources.
The uncertainties of the effective temperature ($T_{\rm eff}$), surface gravity (log$_{g}$), line-of-sight velocity ($V_{\rm r}$),  [$\alpha$/Fe] and [Fe/H], of the whole sample stars, are typically better than, 100\,K, 0.10\,dex, 5\,km s$^{-1}$, 0.03$-$0.05\,dex and 0.10$-$0.15\,dex, respectively \citep{Bovy2014, Huang2020}.
Given the distance from the standard-candle nature of RC stars, their typical accuracies are generally better than 5\%--10\% \citep{Bovy2014, Huang2020}.
Stellar masses and ages are determined from the LAMOST spectra using the kernel principal component analysis (KPCA) method \citep[e.g.,][]{Scholkopf1998}, which is trained with thousands of RC stars in the LAMOST--Kepler fields with accurate asteroseismic mass measurements.
The typical uncertainties of the derived stellar masses and ages are around 15\% and 30\%, respectively.
For the age-dependent analysis presented in this work, the APOGEE RC stars were cross-matched with the LAMOST KPCA age catalog, and only stars with available KPCA age estimates were retained.
This procedure ensures a homogeneous age scale for both the LAMOST and APOGEE samples used in the age analysis, thereby avoiding systematic differences introduced by different age-determination methods.

The standard Galactocentric cylindrical coordinate ($R$, $\phi$, $Z$) has been used for this study, with three velocity components, $V_{R}$, $V_{\phi}$ and $V_{z}$, respectively.
$R$ is the projected Galactocentric distance with increasing radial outwards, $\phi$ is the Galactic disk rotational direction, and $\phi$ = 0$^{\circ}$ points to the Galactic anti-center direction, $Z$ points to the North Galactic Pole.
To transform the heliocentric coordinates to Galactocentric coordinates, we set the local circular velocity as $V_{c,0}$ = 238 km s$^{-1}$ \citep{Reid2004, Schonrich2010, Schonrich2012, Huang2016, Bland-Hawthorn2016}, the Galactocentric distance of the Sun as $R_{\odot}$ = 8.34 kpc \citep{Reid2014}, and solar motions as ($U_{\odot}$, $V_{\odot}$, $W_{\odot}$) $=$ $(13.00, 12.24, 7.24)$ km s$^{-1}$ \citep{Schonrich2018}.

To improve the accuracy of the chemical distribution calculations, we further use cuts with SNRs $>$ 20 and the distance uncertainty $\leq$ 10\%.
To exclude any possibility of the halo stars, we further set stellar [Fe/H] $\geq -1.0$ dex and $\ |V_{z}|$ $\leq$ 120 km s$^{-1}$ \citep{Hayden2020, Sun2020}, and finally, 168,344 RC stars were selected.
The spatial and metallicity distributions of the final selected stars are shown in Fig.\,1.
Since the stellar parameters, such as [Fe/H] and [$\alpha$/Fe], are derived using different analysis pipelines in the APOGEE and LAMOST surveys, a homogenization procedure is performed before combining the datasets. Given that the LAMOST sample is significantly larger than the APOGEE sample, the [Fe/H] and [$\alpha$/Fe] measurements of the APOGEE datasets are calibrated onto the LAMOST datasets based on the best-fitting relations derived from the common targets of the two surveys (see Appendix Fig.\,A1).

Given that the metallicity gradients are distinguished for the thin and thick disk populations \citep[e.g,][]{Yan2019, Sun2024b}, we separate our RC sample stars into thin and thick disks according to their locations on the [Fe/H]--[$\alpha$/Fe] plane (see Appendix Fig.\,B1).
To obtain clean thin/thick disk stars, two empirical cuts are used to separate the two disks \citep[e.g.,][]{Bensby2011, Lee2011, Han2020, Sun2020}, and this chemically based classification has been extensively validated in the literature, where the resulting high-[$\alpha$/Fe] and low-[$\alpha$/Fe] populations exhibit distinct chemical, chemo-kinematic, and chemo-dynamical properties that broadly correspond to the canonical thick and thin disks, respectively \citep[e.g.,][]{Bensby2014, Imig2023, Sun2024a, Sun2024b, Sun2025a}.
Finally, 130,382 stars and 25,199 stars are selected for thin and thick disks, respectively.

To measure radial metallicity gradients, we divide the sample into bins of vertical height and stellar age.
Within each bin, we fit a linear relation between [Fe/H] and Galactocentric radius $R$.
The fitting was performed using the Python package emcee \citep{Foreman-Mackey2013} to generate MCMC samples of the posterior distributions of the fitting parameters.
Assuming Gaussian uncertainties, the negative log-likelihood was defined using the least-squares statistic. We adopted 100 walkers and evolved each chain for 5000 steps, discarding the initial 2000 steps as burn-in.

\begin{figure*}[t]
\begin{center}
\includegraphics[width=16.5cm]{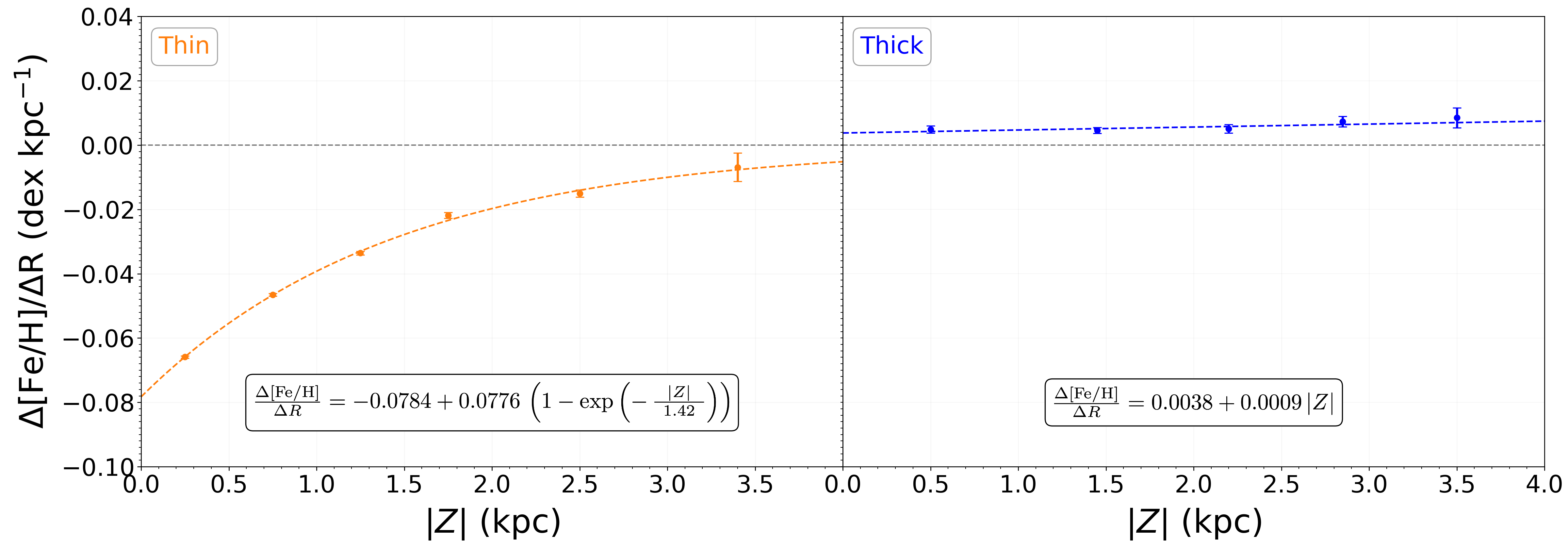}
\end{center}
\begin{center}
\caption{The radial metallicity gradient ($\Delta$[Fe/H]/$\Delta R$) as a function of vertical height ($|Z|$) for thin (left panel) and thick (right panel) disks.
Colored dots represent the measured radial metallicity gradients in individual $|Z|$ bins, with vertical error bars denoting the corresponding 1$\sigma$ uncertainties obtained from the MCMC fits to the [Fe/H]--$R$ relation in each bin.
The dashed curves denote the best-fitting models, parameterized by an exponential saturation function as Equation (1) for the thin disk and a linear relation as Equation (2) for the thick disk.
The best-fitting equations are indicated in each panel.
The number of stars contributing to individual $|Z|$ bins ranges from 523 to 71,413 for the thin disk and from 701 to 9,892 for the thick disk, corresponding to 0.31\%--42.42\% and 0.42\%--5.88\% of the total sample, respectively.}
\end{center}
\end{figure*}
%%\label{fig2}

\section{Results and Discussion}

The radial metallicity gradients ($\Delta$[Fe/H]/$\Delta$R) as a function of vertical height ($|Z|$) for the two disks are shown in Fig.\,2.

The thin disk exhibits a pronounced negative $\Delta$[Fe/H]/$\Delta$R near the Galactic mid-plane that systematically flattens with increasing $|Z|$, following an approximately exponential flattening trend from $-$0.0784\,dex\,kpc\,$^{-1}$ at $|Z|$ $\sim$ 0.0\,kpc to $-$0.0075\,dex\,kpc\,$^{-1}$ at $|Z|$ $\sim$ 3.5\,kpc (see right panel of Fig.\,2).
The thick disk displays an almost constant week positive $\Delta$[Fe/H]/$\Delta$R (nearly $+$0.0050 $\sim$ $+$0.0080\,dex\,kpc\,$^{-1}$) over Galactic height (see right panel of Fig.\,2).

Our results indicate that the thin disk exhibits $\Delta$[Fe/H]/$\Delta$R of $-$0.0658 $\pm$ 0.0004 dex\,kpc$^{-1}$ at $|Z|$ $\sim$ 0.25\,kpc, and the thick disk shows weak positive $\Delta$[Fe/H]/$\Delta$R of $+$0.0050 $\pm$ 0.0014 dex\,kpc$^{-1}$ at $|Z|$ $\sim$ 2.25 kpc (see Fig.\,2), which are in good agreement with the A/F/G/K-type giant sample of Yan et al. ({\color{blue}{2019}}).
The thick disk displays a globally weakly positive $\Delta$[Fe/H]/$\Delta$R \citep[this positive gradient has also been observed in other works; e.g.,][]{Vickers2021, Imig2023}, and it shows no obvious change with $|Z|$, which is slightly different from the result of Yan et al. ({\color{blue}{2019}}), while it is broadly consistent with the results of old/high-$\alpha$ populations determined by the MSTO-SG sample of Wang et al. ({\color{blue}{2019}}), the RC sample of Sun et al. ({\color{blue}{2024a}}), and the red giant sample of Imig et al. ({\color{blue}{2023}}).
Since the distances of giants are generally obtained from Gaia parallaxes, the discrepancy between our results and theirs is likely due to their larger distance uncertainties compared to the more precise RC distances adopted in this work.

To intuitively quantify the variation of radial metallicity gradient ($\nabla_{\rm R}=\Delta$[Fe/H]/$\Delta$R) as a function of $|Z|$ for the two disks, we fit the global trend of the thin disk with a saturating exponential function, as follows:

\begin{equation}
\label{eq:tiltangle2}
    \nabla_{\mathrm{R}} (|Z|) = \nabla_{\mathrm{R},0} + \Delta\nabla_{\mathrm{R}}\,\left(1 - \exp\left(-\,\frac{|Z|}{Z_{\mathrm{R}}}\right)\right)
\end{equation}
and fit the relation of the thick disk with a linear function follows:
\begin{equation}
\label{eq:tiltangle2}
    \nabla_{\mathrm{R}} (|Z|) = \nabla_{\mathrm{R},0} + k_{\mathrm{R}}\,|Z|
\end{equation}
Here, $\nabla_{\mathrm{R},0}$ represents the radial metallicity gradient at the Galactic mid-plane ($|Z| = 0.0$\,kpc).
For thin disk, $Z_{\mathrm{R}}$ represents the characteristic scale height of the exponential flattening of $\nabla_{\mathrm{R}}$, $\Delta\nabla_{\mathrm{R}}$ represents exponential saturation amplitude of $\nabla_{\mathrm{R}}$ with $|Z|$.
For thick disk, $k_{\mathrm{R}}$ represents the linear variation rate of $\nabla_{\mathrm{R}}$ with $|Z|$.

The global trends of the $\Delta$[Fe/H]/$\Delta$R--$|Z|$ of the thin and thick disks are fitted by the dotted line in Fig.\,2, and the corner plots of the posterior distributions of MCMC samples of the parameters are presented in Appendix Fig.\,C1.
The results yield the thin and thick disks are, respectively, $\Delta \mathrm{[Fe/H]}/\Delta R$ = $-$0.0784 $+$ 0.0776\,(1 $-$ exp\,($-$\,$|Z|$\,/\,1.42)) and $\Delta \mathrm{[Fe/H]}/\Delta R$  = 0.0038 $+$ 0.0009\,$|Z|$.

The thin disk exhibits a pronounced negative $\Delta$[Fe/H]/$\Delta$R (around $-$0.0784 dex\,kpc$^{-1}$) near the Galactic mid-plane ($|Z|$ = 0.0 kpc) that progressively flattens with increasing vertical height, reaching an almost flat value of around $-$0.0008\,dex\,kpc$^{-1}$ at the highest $|Z|$.
This vertical dependence suggests a close coupling between the chemical and dynamical evolution of the thin disk, since stars at larger vertical heights are typically kinematically hotter and therefore more strongly affected by radial migration \citep{Yan2019, Hackshaw2024, Sun2024b}. Such enhanced radial mixing can naturally flatten the radial metallicity gradient at larger $|Z|$.

The thick disk shows weakly positive $\Delta$[Fe/H]/$\Delta$R that remains nearly invariant with increasing $|Z|$, with a slope of $k_{\mathrm{R}}$ = 0.0009$^{+0.0006}_{-0.0007}$\,dex\,kpc$^{-2}$, indicating no significant vertical dependence.
This behavior is likely to arise from multiple viable formation channels, including secular evolution processes such as ``inside-out" and ``upside-down" star formation accompanied by long-term stellar migration \citep[e.g.,][]{Kawata2017, Schonrich2017, Bird2021}, as well as a formation environment characterized by mergers of gas-rich systems and/or the turbulent interstellar medium \citep[ISM; e.g.,][]{Brook2012, Wisnioski2015}.

\begin{figure*}[t]
\begin{center}
\includegraphics[width=16.5cm]{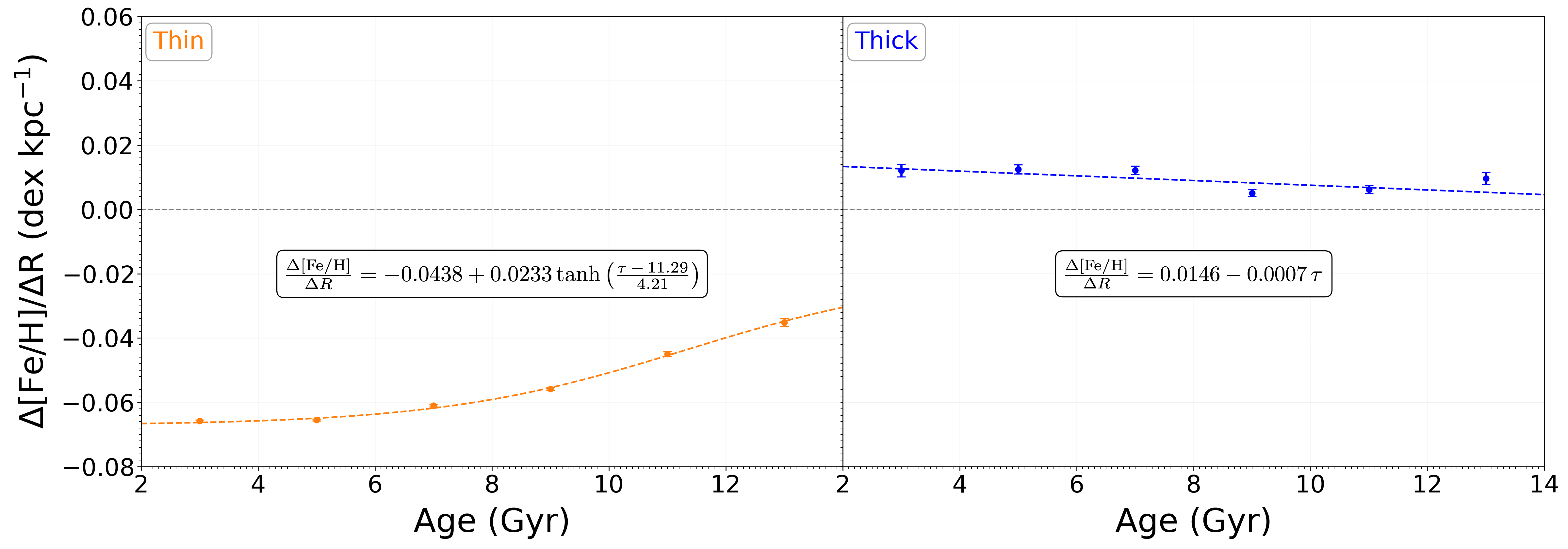}
\end{center}

\begin{center}
\caption{The $\Delta$[Fe/H]$/\Delta$R as a function of stellar age ($\tau$) for thin (left panel) and thick (right panel) disks.
Colored dots indicate the measured radial metallicity gradients in individual age bins, with vertical error bars representing the corresponding 1$\sigma$ uncertainties derived from the MCMC analysis.
The dashed curves show the best-fitting age-dependent models, described by a hyperbolic tangent function for the thin disk and a linear relation for the thick disk.
The best-fitting equations are displayed in each panel.
The number of stars contributing to individual age bins ranges from 1,329 to 46,159 for the thin disk and from 1,885 to 6,949 for the thick disk, corresponding to 0.79\%--27.42\% and 1.12\%--4.13\% of the total sample, respectively.}

\end{center}

\end{figure*}
%%\label{fig3}

To further constrain the origin of the distinct $\Delta$[Fe/H]$/\Delta R$ trends in the two disks, we examine their age dependence.
Fig.\,3 presents the $\Delta$[Fe/H]$/\Delta R$ as a function of stellar age for the thin and thick disks.
The best fitting curve as shown in the figure, and the corner plots of the posterior
distribution of the MCMC samples of the parameters are displayed in Appendix Fig.\,C2.
The thin disk exhibits a clear age dependence in $\Delta$[Fe/H]$/\Delta R$, evolving smoothly from a strong gradient regime of $\Delta$[Fe/H]$/\Delta R$ = $-$0.0445 dex\,kpc$^{-1}$ for young stars (age $\lesssim$ 4.0\,Gyr) to a weak gradient regime of $\Delta$[Fe/H]$/\Delta R$ = $-$0.0220\,dex\,kpc$^{-1}$ for old stars (age $\gtrsim$ 13\,Gyr), which can be well described by 
\begin{equation}
\nabla_{\mathrm{R}} (\tau) = -0.0438 + 0.0233\,\tanh\left(\frac{\tau-11.29}{4.21}\right)
\end{equation}
indicating a gradual time evolution of the radial metallicity gradient for the thin disk.
In contrast, the thick disk exhibits a statistically constant $\Delta$[Fe/H]$/\Delta R$ (nearly $+$0.0065 $\sim$ $+$0.0135\,dex\,kpc\,$^{-1}$) and remains nearly invariant across the full age range, which is approximately described by 
\begin{equation}
\nabla_{\mathrm{R}} (\tau) = 0.0146 - 0.0007\,\tau
\end{equation}
with a slope of $-$0.0007$^{+0.0002}_{-0.0002}$ dex\,kpc$^{-1}$\,Gyr$^{-1}$, indicating no significant evolution with stellar age.
In the context of the theoretical predictions for the temporal evolution of radial metallicity gradients in disk galaxies \citep{Pilkington2012}, the observed contrasting trends in $\Delta$[Fe/H]$/\Delta R$ as a function of stellar age for the thin and thick disks suggest that these two disks likely experienced fundamentally different formation and evolutionary histories.

These results indicate that the thin disk exhibits a clear evolution of the $\Delta$[Fe/H]$/\Delta R$ with stellar age, consistent with a relatively ordered phase of disk evolution, during which radial metallicity gradients were gradually established and maintained through ``inside-out" disk growth \citep[e.g.,][]{Chiappini2009, Frankel2019} and long-term secular evolution processes, particularly stellar radial migration \citep[e.g.,][]{Sellwood2002, Schonrich2009, Minchev2013, Sun2025a}.
This interpretation is further supported by the observed $\Delta$[Fe/H]$/\Delta R$–$|Z|$ relation (Fig.\,2).
The thick disk displays temporally invariant $\Delta$[Fe/H]$/\Delta R$, implying that its radial metallicity gradient was established during an early phase when the disk was dynamically hot and efficiently mixed and has remained unchanged over time.
This lack of temporal evolution disfavors secular evolution scenarios involving ``inside-out" and ``upside-down" star formation accompanied by long-term stellar migration.
Our results are therefore more consistent with the scenario in which thick disk stars formed in an environment characterized by mergers of gas-rich systems and/or the turbulent ISM \citep{Brook2012, Wisnioski2015, Sun2024a}.

It is worth emphasizing that precise stellar age determinations remain challenging \citep{Soderblom2010}, and the associated uncertainties may affect the measurements of $\Delta$[Fe/H]$/\Delta R$ as a function of stellar age, thereby limiting constraints on the detailed evolutionary history of the Galactic disk \citep[e.g.,][]{Sharma2021, Sun2025a}.
Future measurements with improved age precision will provide stronger constraints on the temporal evolution of radial metallicity gradients and the assembly history of the Milky Way disk.

\section{Conclusions}

Using a sample of 200,388 RC stars selected from APOGEE and LAMOST, we investigated the radial metallicity gradients of the Galactic disk as functions of vertical height and stellar age for thin and thick disks. We find that:

The thin disk exhibits a pronounced negative $\Delta$[Fe/H]$/\Delta R$ near the Galactic mid-plane that progressively flattens with increasing $|Z|$, increasing steadily from $-$0.0784\,dex\,kpc\,$^{-1}$ at $|Z|$ $\sim$ 0.0\,kpc to $-$0.0075\,dex\,kpc\,$^{-1}$ at $|Z|$ $\sim$ 3.5\,kpc, following $\Delta \mathrm{[Fe/H]}/\Delta R$ = $-$0.0784 $+$ 0.0776\,(1 $-$ exp($-$\,$|Z|$/1.42)).
The thin disk also exhibits a clear age dependence in $\Delta$[Fe/H]$/\Delta R$, evolving smoothly from a strong gradient regime of $-$0.0445 dex\,kpc$^{-1}$ for young stars (age $<$ $\sim$4.0\,Gyr) to a weak-gradient regime of $-$0.0220\,dex\,kpc$^{-1}$ for old stars (age $>$ $\sim$13\,Gyr), following $\Delta \mathrm{[Fe/H]}/\Delta R$ = $-$0.0438 $+$ 0.0233\,tanh\,(($\tau$\,$-$\,11.29)\,/\,4.21).

The thick disk shows weakly positive radial metallicity gradients that remain statistically invariant with respect to both vertical height and stellar age, following $\Delta \mathrm{[Fe/H]}/\Delta R$  = 0.0038 $+$ 0.0009\,$|Z|$ and $\Delta \mathrm{[Fe/H]}/\Delta R$  = 0.0146 $-$ 0.0007\,$\tau$, respectively.

These results indicate that the thin disk retains radial metallicity gradients shaped by relatively ordered inside-out growth and long-term secular evolution processes, including stellar radial migration.
The thick disk exhibits spatially and temporally homogeneous radial metallicity gradients, which are consistent with a formation environment characterized by mergers of gas-rich systems and/or the turbulent ISM.

\section*{Acknowledgements}

We thank the anonymous referee for very the useful suggestions to improve the work.
This work is supported by the NSFC projects 12133002, 11833006, and 11811530289, and the National Key R\&D Program of China No. 2019YFA0405500, 2019YFA0405503, and CMS-CSST-2021-A09, and the CPSF under Grant Number GZC20240125, 2024M760240, 2025T180873 and YJA20250180, and the Tsinghua Shuimu Scholar Number 2025SM415.

Guoshoujing Telescope (the Large Sky Area Multi-Object Fiber Spectroscopic Telescope LAMOST) is a National Major Scientific Project built by the Chinese Academy of Sciences. Funding for the project has been provided by the National Development and Reform Commission. LAMOST is operated and managed by the National Astronomical Observatories, Chinese Academy of Sciences.

\bibliographystyle{aasjournal}

\begin{thebibliography}{}

\bibitem[Andrews et al.(2017)]{Andrews2017} Andrews, B.~H., Weinberg, D.~H., Sch{\"o}nrich, R., et al.\ 2017, \apj, 835, 224
\bibitem[Andrievsky et al.(2002)]{Andrievsky2002} Andrievsky, S.~M., Kovtyukh, V.~V., Luck, R.~E., et al.\ 2002, \aap, 381, 32
\bibitem[Andrievsky et al.(2004)]{Andrievsky2004} Andrievsky, S.~M., Luck, R.~E., Martin, P., et al.\ 2004, \aap, 413, 159
\bibitem[Balser et al.(2011)]{Balser2011} Balser, D.~S., Rood, R.~T., Bania, T.~M., et al.\ 2011, \apj, 738, 1, 27
\bibitem[Bensby et al.(2011)]{Bensby2011} Bensby, T., Alves-Brito, A., Oey, M. S., Yong, D., Mel{\'e}ndez, J.,\ 2011, \apjl, 735, L46
\bibitem[Bensby, Feltzing \& Oey(2014)]{Bensby2014} Bensby, T., Feltzing, S., \& Oey, M. S.\ 2014, \aap, 562, A71
\bibitem[Bird et al.(2021)]{Bird2021} Bird, J.~C., Loebman, S.~R., Weinberg, D.~H., et al.\ 2021, \mnras, 503, 1815
\bibitem[Bland-Hawthorn \& Gerhard(2016)]{Bland-Hawthorn2016} Bland Hawthorn, J., Gerhard, O. 2016, ARA\&A, 54, 529
\bibitem[Boeche et al.(2013)]{Boeche2013} Boeche, C., Siebert, A., Piffl, T., et al.\ 2013, \aap, 559, A59
\bibitem[Boeche et al.(2014)]{Boeche2014} Boeche, C., Siebert, A., Piffl, T., et al.\ 2014, \aap, 568, A71
\bibitem[Bovy et al.(2014)]{Bovy2014} Bovy, J., Nidever, D. L., Rix, H. W., et al.\ 2014, \apj, 790, 127
\bibitem[Brook et al.(2012)]{Brook2012} Brook, C. B., Stinson, G. S., Gibson, B. K., et al.\ 2012, \mnras, 426, 690
\bibitem[Chen et al.(2003)]{Chen2003} Chen, L., Hou, J.~L., \& Wang, J.~J.\ 2003, \aj, 125, 3, 1397
\bibitem[Chiappini(2009)]{Chiappini2009} Chiappini, C. 2009, in IAU Symp. 254, The Galaxy Disk in Cosmological Context, ed. Andersen, J., Bland-Hawthorn, J., \& Nordstr{\"o}m, B., (Cambridge: Cambridge Univ. Press), 191
\bibitem[Costa et al.(2004)]{Costa2004} Costa, R.~D.~D., Uchida, M.~M.~M., \& Maciel, W.~J.\ 2004, \aap, 423, 199
\bibitem[Cui et al.(2012)]{Cui2012} Cui, X. Q., Zhao, Y. H., Chu, Y. Q., et al.\ 2012,\ RAA, 12, 1197
\bibitem[Daflon \& Cunha(2004)]{Daflon2004} Daflon, S. \& Cunha, K.\ 2004, \apj, 617, 2, 1115
\bibitem[Deng et al.(2012)]{Deng2012} Deng, L. C., Newberg, H. J., Liu, C., et al.\ 2012,\ RAA, 12, 735
\bibitem[Foreman-Mackey et al.(2013)]{Foreman-Mackey2013} Foreman-Mackey, D., Hogg, D. W., Lang, D., et al.\ 2013, PASP, 125, 306
\bibitem[Frankel et al.(2019)]{Frankel2019} Frankel, N., Sanders, J., Rix, H.-W., et al.\ 2019, \apj, 884, 99
\bibitem[Grisoni et al.(2018)]{Grisoni2018} Grisoni, V., Spitoni, E., \& Matteucci, F.\ 2018, \mnras, 481, 2570
\bibitem[Han et al.(2020)]{Han2020} Han, D. R., Lee, Y. S., Kim, Y. K., et al.\ 2020, \apj, 896, 14H
\bibitem[Hackshaw et al.(2024)]{Hackshaw2024} Hackshaw, Z., Hawkins, K., Filion, C., et al.\ 2024, \apj, 977, 2, 143
\bibitem[Hawkins(2023)]{Hawkins2023} Hawkins, K.\ 2023, \mnras, 525, 3, 3318
\bibitem[Hayden et al.(2014)]{Hayden2014} Hayden, M.~R., Holtzman, J.~A., Bovy, J., et al.\ 2014, \aj, 147, 5, 116
\bibitem[Hayden et al.(2020)]{Hayden2020} Hayden, M. R., Bland-Hawthorn, J., Sharma, S., et al.,\ 2020, \mnras, 493, 2952
\bibitem[Henry et al.(2010)]{Henry2010} Henry, R.~B.~C., Kwitter, K.~B., Jaskot, A.~E., et al.\ 2010, \apj, 724, 1, 748
\bibitem[Huang et al.(2015)]{Huang2015} Huang, Y., Liu X. W., Yuan H. B., et al.,\ 2015, \mnras, 449, 162
\bibitem[Huang et al.(2016)]{Huang2016} Huang, Y., Liu, X. W., Yuan, H. B., et al.,\ 2016, \mnras, 463, 2623
\bibitem[Huang et al.(2020)]{Huang2020} Huang, Y., Sch{\"o}nrich, R., Zhang, H. W., et al., 2020, ApJS, 249, 29
\bibitem[Imig et al.(2023)]{Imig2023} Imig, J., Price, C., Holtzman, J.~A., et al.\ 2023, \apj, 954, 124
\bibitem[Katz et al.(2011)]{Katz2011} Katz, D., Soubiran, C., Cayrel, R., et al.\ 2011, \aap, 525, A90
\bibitem[Kawata et al.(2017)]{Kawata2017} Kawata, D., Grand, R. J. J., Gibson, B. K., et al.\ 2017, \mnras, 464, 702
\bibitem[Larson(1976)]{Larson1976} Larson, R.~B.\ 1976, \mnras, 176, 31
\bibitem[Lee et al.(2011)]{Lee2011} Lee, Y. S., Beers, T. C., An, D., et al.\ 2011, \apj, 738, 187
\bibitem[Luck et al.(2006)]{Luck2006} Luck, R.~E., Kovtyukh, V.~V., \& Andrievsky, S.~M.\ 2006, \aj, 132, 2, 902
\bibitem[Mackereth et al.(2019)]{Mackereth2019} Mackereth, J. T., Bovy, J., Leung, H. W., et al.\ 2019, \mnras, 489, 176
\bibitem[Magrini et al.(2009)]{Magrini2009} Magrini, L., Sestito, P., Randich, S., et al.\ 2009, \aap, 494, 1, 95
\bibitem[Majewski et al.(2017)]{Majewski2017} Majewski, S. R., Schiavon, R. P., Frinchaboy, P. M., et al.\ 2017, \aj, 154, 94
\bibitem[Matteucci \& Francois(1989)]{Matteucci1989} Matteucci, F. \& Francois, P.\ 1989, \mnras, 239, 885
\bibitem[Matteucci(2012)]{Matteucci2012} Matteucci, F.,\ 2012,\ Chemical Evolution of Galaxies (Berlin: Springer)
\bibitem[Minchev et al.(2013)]{Minchev2013} Minchev, I., Chiappini, C., \& Martig, M.\ 2013, \aap, 558, A9

\bibitem[Pilkington et al.(2012)]{Pilkington2012} Pilkington, K., Few, C.~G., Gibson, B.~K., et al.\ 2012, \aap, 540, A56.

\bibitem[Pezzulli \& Fraternali(2016)]{Pezzulli2016} Pezzulli, G. \& Fraternali, F.\ 2016, \mnras, 455, 2308
\bibitem[Quinn et al.(1993)]{Quinn1993} Quinn, P. J., Hernquist, L., \& Fullagar, D. P.\ 1993, \apj, 403, 74
\bibitem[Reid \& Brunthaler(2004)]{Reid2004} Reid, M. J., \& Brunthaler, A.\ 2004, \apj, 616, 872
\bibitem[Reid et al.(2014)]{Reid2014} Reid, M. J., Menten, K. M., Brunthaler, A., et al.\ 2014, \apj, 783, 130
\bibitem[Scarano \& L{\'e}pine(2013)]{Scarano2013} Scarano, S. \& L{\'e}pine, J.~R.~D.\ 2013, \mnras, 428, 625
\bibitem[Sch{\"o}lkopf et al.(1998)]{Scholkopf1998} Sch{\"o}lkopf, B., Smola, A.~J., \& M{\"u}ller, K.-R. 1998, Neural Computation, 10, 1299
\bibitem[Sch{\"o}nrich \& Binney(2009)]{Schonrich2009} Sch{\"o}nrich, R., Binney, J.\ 2009, \mnras, 399, 1145
\bibitem[Sch{\"o}nrich et al.(2010)]{Schonrich2010} Sch{\"o}nrich, R., Binney, J., \& Dehnen, W.\ 2010, \mnras, 403, 1829
\bibitem[Sch{\"o}nrich et al.(2012)]{Schonrich2012} Sch{\"o}nrich, R. \ 2012, \mnras, 427, 274
\bibitem[Sch{\"o}nrich \& McMillan(2017)]{Schonrich2017} Sch{\"o}nrich, R., \& McMillan, P. J.\ 2017, \mnras, 467, 1154
\bibitem[Sch{\"o}nrich \& Dehnen(2018)]{Schonrich2018} Sch{\"o}nrich, R., \& Dehnen, W.\ 2018, \mnras, 478, 3809
\bibitem[Sellwood \& Binney(2002)]{Sellwood2002} Sellwood, J.~A. \& Binney, J.~J.\ 2002, \mnras, 336, 3, 785
\bibitem[Soderblom (2010)]{Soderblom2010} Soderblom, D.,\ 2010, ARA\&A, 48, 581
\bibitem[Sharma et al.(2021)]{Sharma2021} Sharma, S., Hayden, M. R., Bland-Hawthorn, J., et al.\ 2021, \mnras, 506, 1761
\bibitem[Sun et al.(2020)]{Sun2020} Sun, W. X., Huang, Y., Wang, H. F., et al.\ 2020, \apj, 903, 12
\bibitem[Sun et al.(2024a)]{Sun2024a} Sun, W. X., Shen, H., Jiang, B. W., \& Liu, X. W. .\ 2024a, \apjs, 272, 8
\bibitem[Sun et al.(2024b)]{Sun2024b} Sun, W. X., Huang, Y., Shen, H., et al.\ 2024b, \apj, 961, 141
\bibitem[Sun et al.(2025)]{Sun2025a} Sun, W. X., Huang, Y., Shen, H., et al.\ 2025, \apj, 979, 103
\bibitem[Vickers et al.(2021)]{Vickers2021} Vickers, J. J., Shen, J. T., Li, Z. Y.\ 2021, \apj, 922, 189
\bibitem[Wang et al.(2019)]{Wang2019} Wang, C., Liu, X.-W., Xiang, M.-S., et al.\ 2019, \mnras, 482, 2189
\bibitem[Wang \& Chen(2021)]{Wang2021} Wang, S. \& Chen, X.\ 2021, \apj, 923, 2, 145.
%\bibitem[Wang et al.(2023)]{Wang2023} Wang, C., Huang, Y., Zhou, Y., et al. \ 2023, \aap, 675, A26
\bibitem[Wisnioski et al.(2015)]{Wisnioski2015} Wisnioski, E., F{\"o}rster Schreiber, N. M., Wuyts, S., et al.\ 2015, \apj, 799, 209
\bibitem[Xiang et al.(2015)]{Xiang2015} Xiang, M. S., Liu, X. W., Yuan, H. B., et al.\ 2015, \mnras, 448, 822
\bibitem[Yan et al.(2019)]{Yan2019} Yan, Y., Du, C., Liu, S., et al.\ 2019, \apj, 880, 36

\end{thebibliography}

\appendix

\section{Comparison and calibration of [Fe/H] and [$\alpha$/Fe] between the APOGEE and LAMOST datasets}

This appendix presents the comparison and calibration of [Fe/H] (left) and [$\alpha$/Fe] (right) between the APOGEE and LAMOST datasets (see Fig.\,{\color{blue}{A1}}).

\begin{figure*}[t]

\begin{center}
\subfigure{
\includegraphics[width=8.cm]{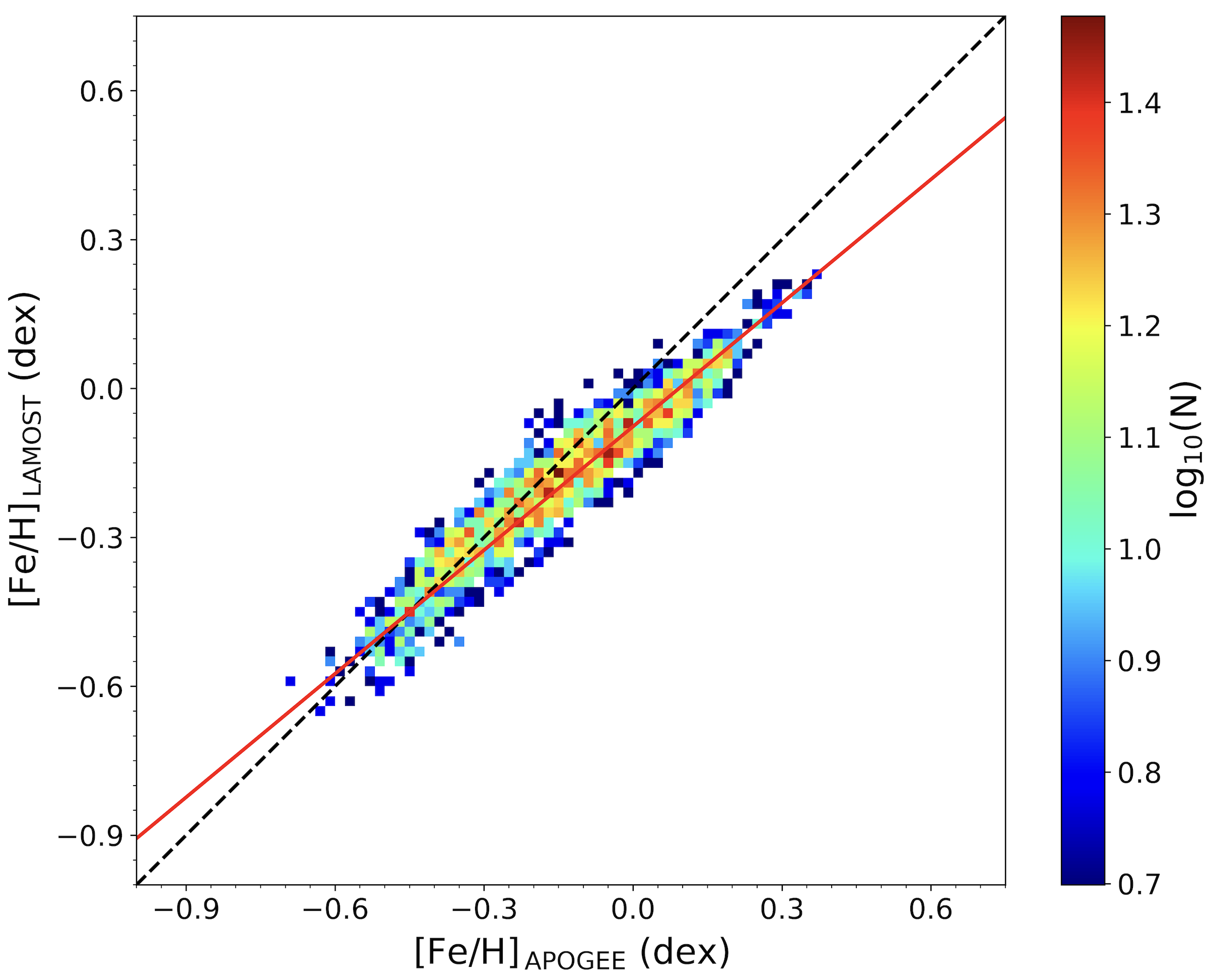}
}
\hspace{0.9cm}
\subfigure{
\includegraphics[width=8.cm]{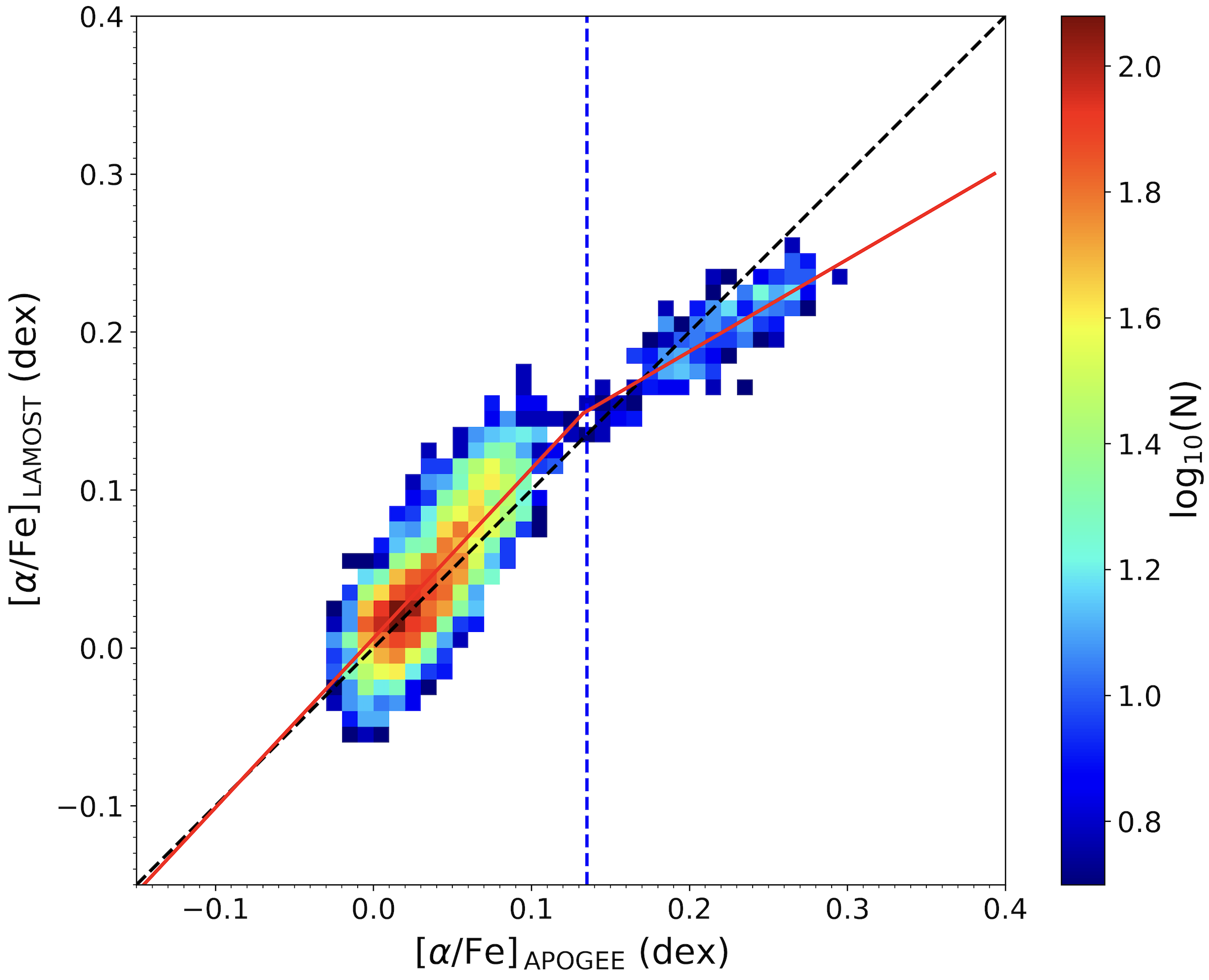}
}
\end{center}
\caption{Comparison and calibration of [Fe/H] (left) and [$\alpha$/Fe] (right) between the APOGEE and LAMOST datasets based on common targets between the two surveys.
Black dashed lines indicate one-to-one relations, and the red solid lines show the best-fitting linear calibration relations to the data.
Each bin contains at least five stars, with bin sizes of 0.02,dex in both [Fe/H] and [$\alpha$/Fe].}

\end{figure*}
%%\label{figA1}

\section{Separation of the thin and thick disks in the [Fe/H]--[$\alpha$/Fe] plane}

This appendix displays the separation of the thin and thick disks in the [Fe/H]--[$\alpha$/Fe] plane (see Fig.\,{\color{blue}{B1}}).

\begin{figure}[t]
\begin{center}
\includegraphics[width=8.5cm]{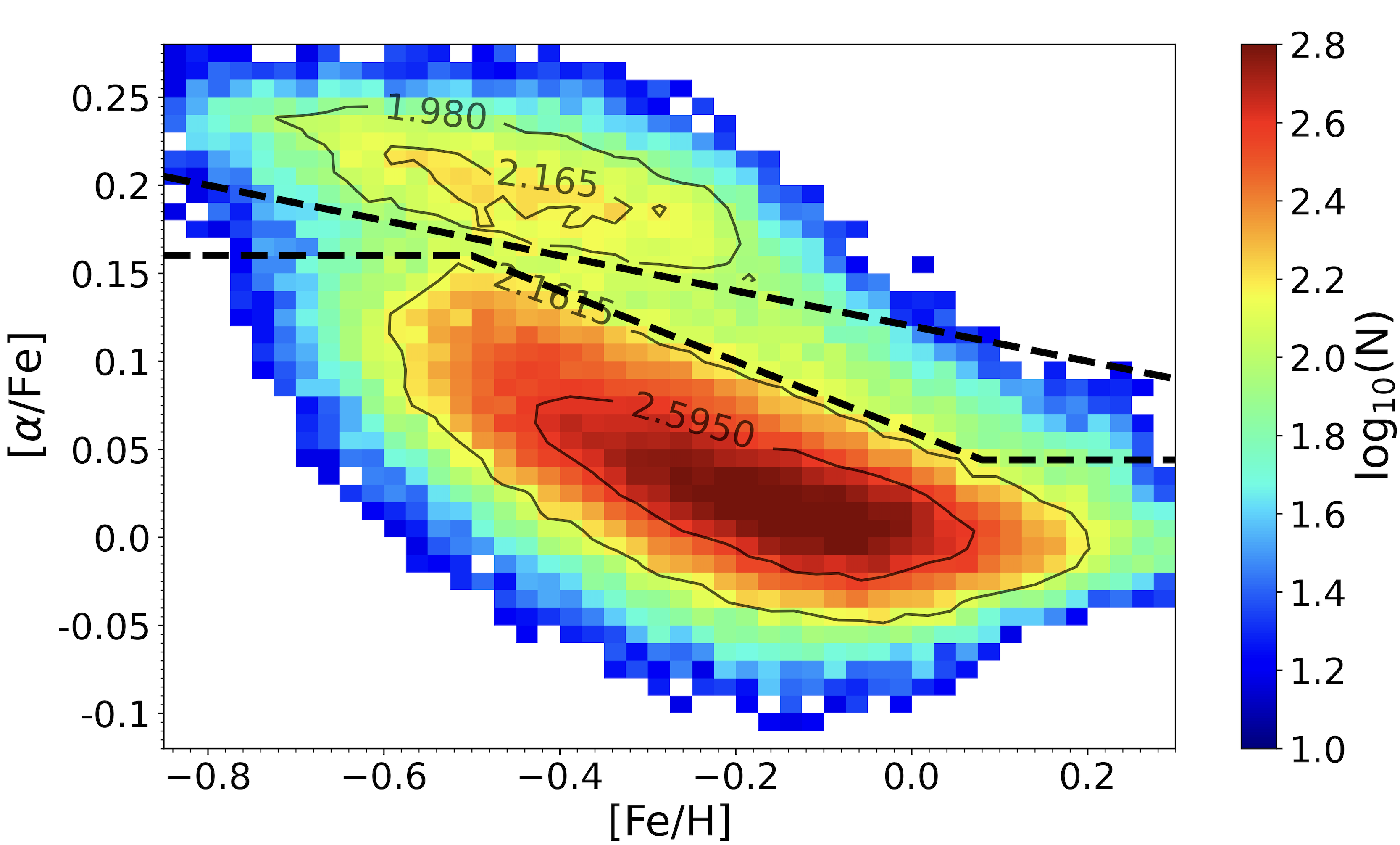}
\caption{Distribution of the RC sample in the [Fe/H]$-$[$\alpha$/Fe] plane. The color scale represents the logarithmic stellar number density, and the contours indicate isodensity levels. The density map is constructed using bin sizes of 0.025 dex in [Fe/H] and 0.02 dex in [$\alpha$/Fe], with only bins containing at least 15 stars included.
The two dashed lines are used to separate the thick (above the lines) and the thin (below the lines) disk stars.}
\end{center}
\end{figure}
%%\label{figB1}

\section{Example MCMC posterior distributions for the parameters of the thin and thick disk models}

Fig.\,{\color{blue}{C1}} shows example MCMC posterior distributions of the fitted parameters for the thin disk model described by Equation (1) (left panel) and the thick disk model described by Equation (2) (right panel).

Fig.\,{\color{blue}{C2}} displays example MCMC posterior distributions of the fitted parameters for the age-dependent models of the thin (left panel) and thick (right panel) disks.

\begin{figure*}[t]
\centering

\subfigure{
\includegraphics[width=8.3cm]{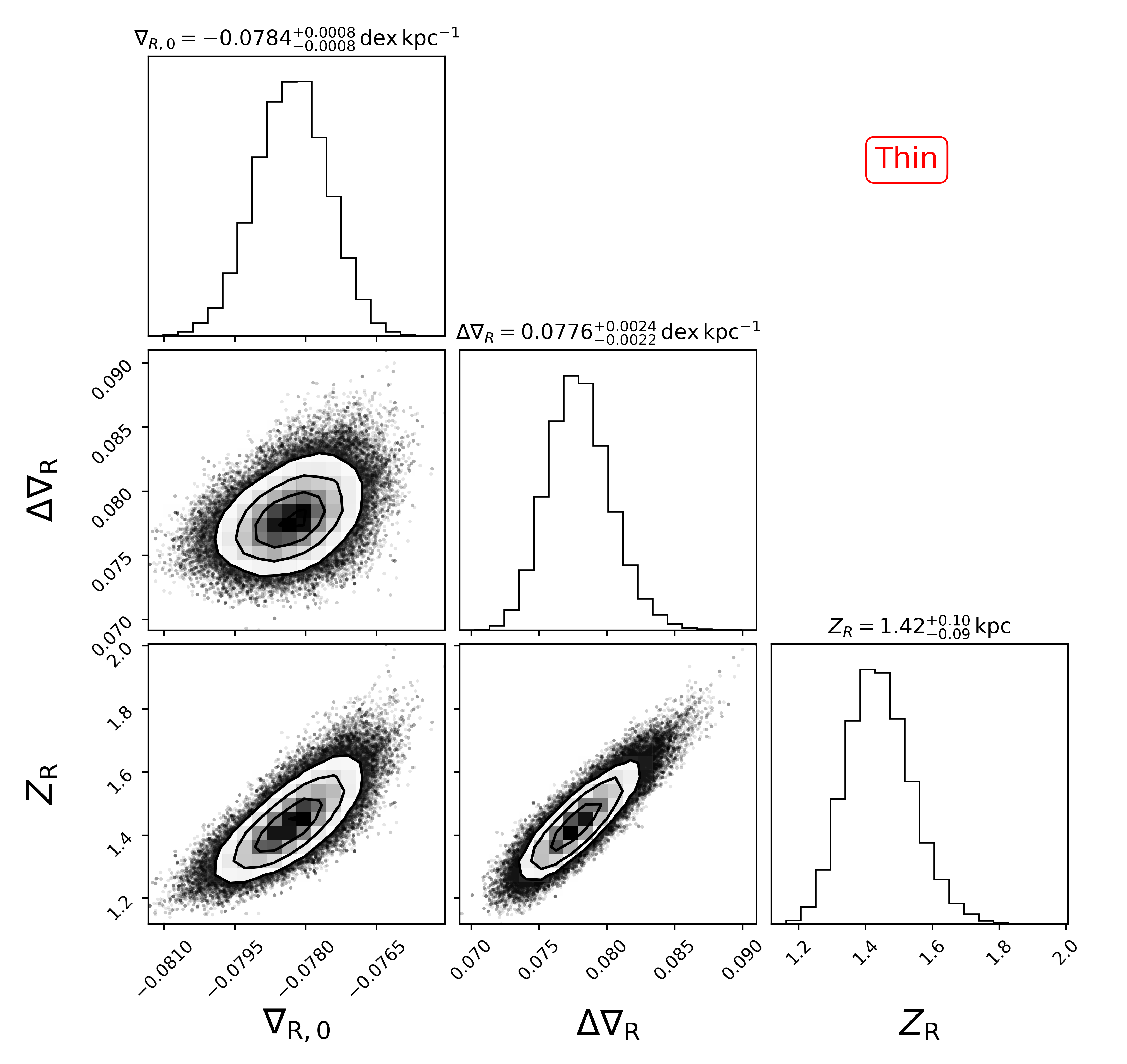}
}
\hspace{0.5cm}
\subfigure{
\includegraphics[width=8.3cm]{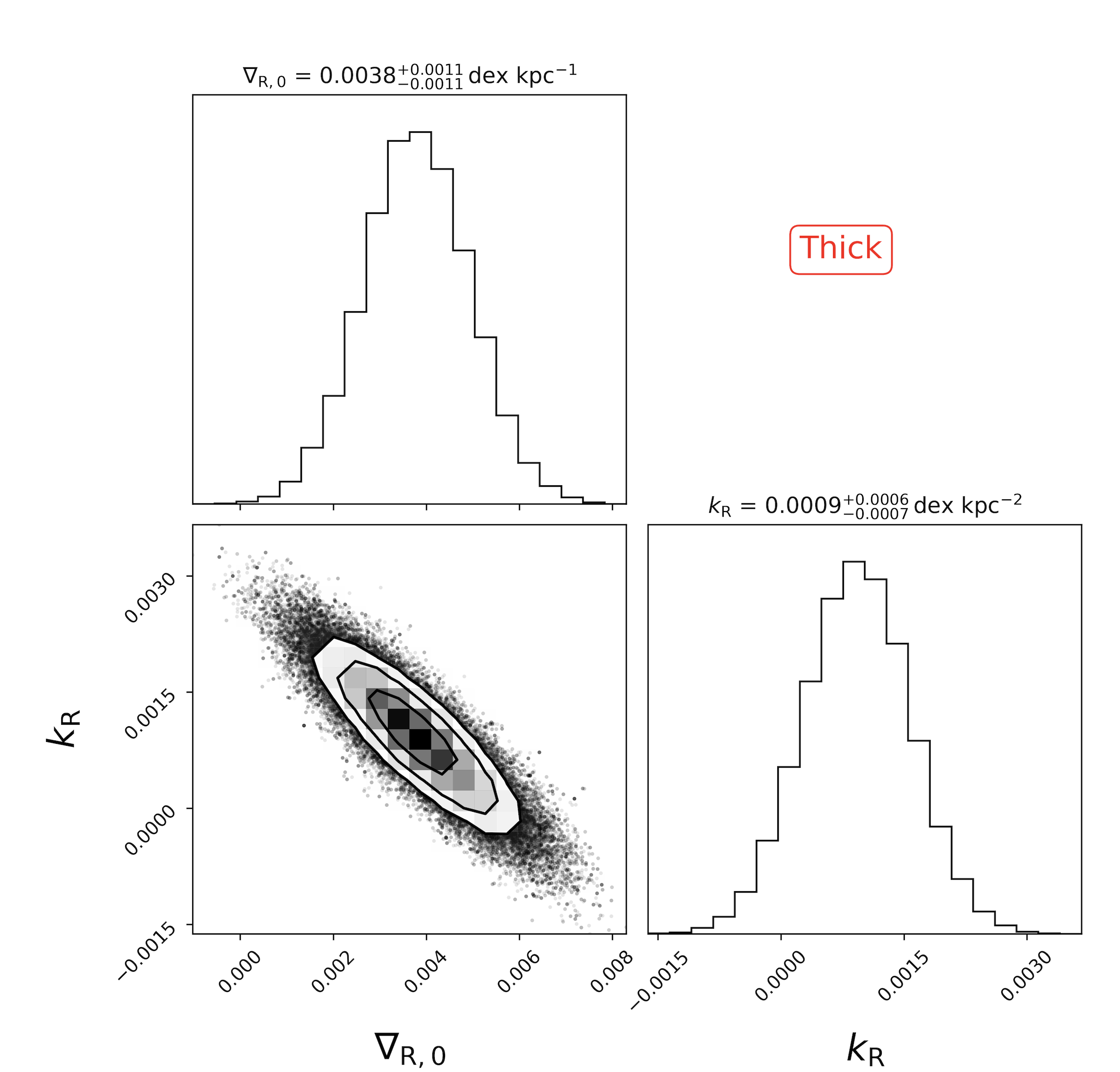}
}

\caption{Corner plots of the posterior distributions of the MCMC samples for the parameters of the thin disk (left panel) and thick disk (right panel) models.
The thin disk is fitted with $\nabla_{\mathrm{R}}\,(|Z|)$ = $\nabla_{\mathrm{R},0}$ $+$ $\Delta\nabla_{\mathrm{R}}$\,$\left(1-\exp\left(-|Z|/Z_{\mathrm{R}}\right)\right)$, whereas the thick disk is fitted with $\nabla_{\mathrm{R}}\,(|Z|)$ = $\nabla_{\mathrm{R},0}$ $+$ $k_{\mathrm{R}}$\,$|Z|$.
The shadowed contours from inside to outside correspond to the confidence intervals of 1$\sigma$, 2$\sigma$ and 3$\sigma$, respectively.}
\end{figure*}
%%\label{Fig.C1}

\begin{figure*}[t]
\centering

\subfigure{
\includegraphics[width=8.6cm]{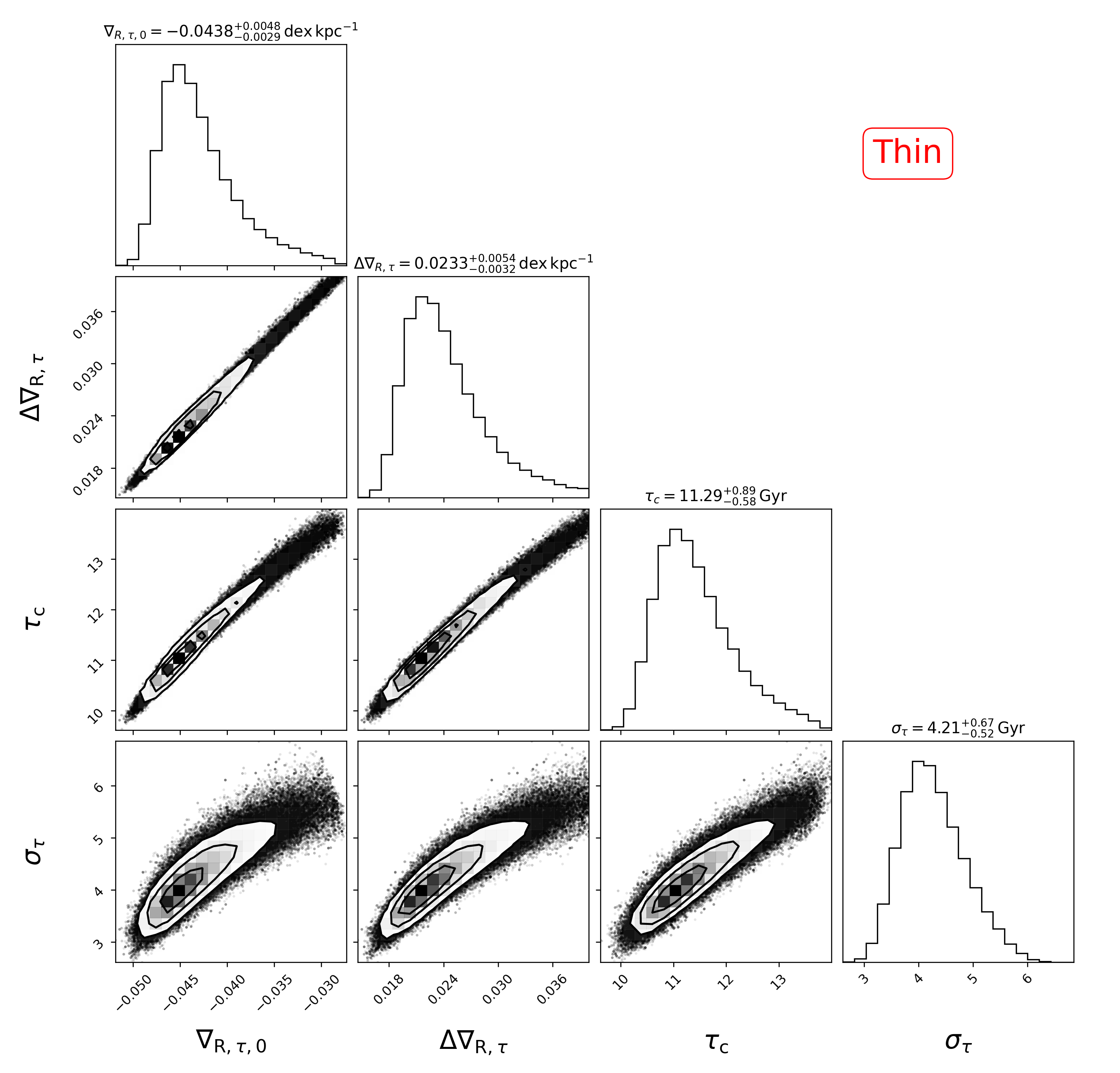}
}
\hspace{0.2cm}
\subfigure{
\includegraphics[width=8.6cm]{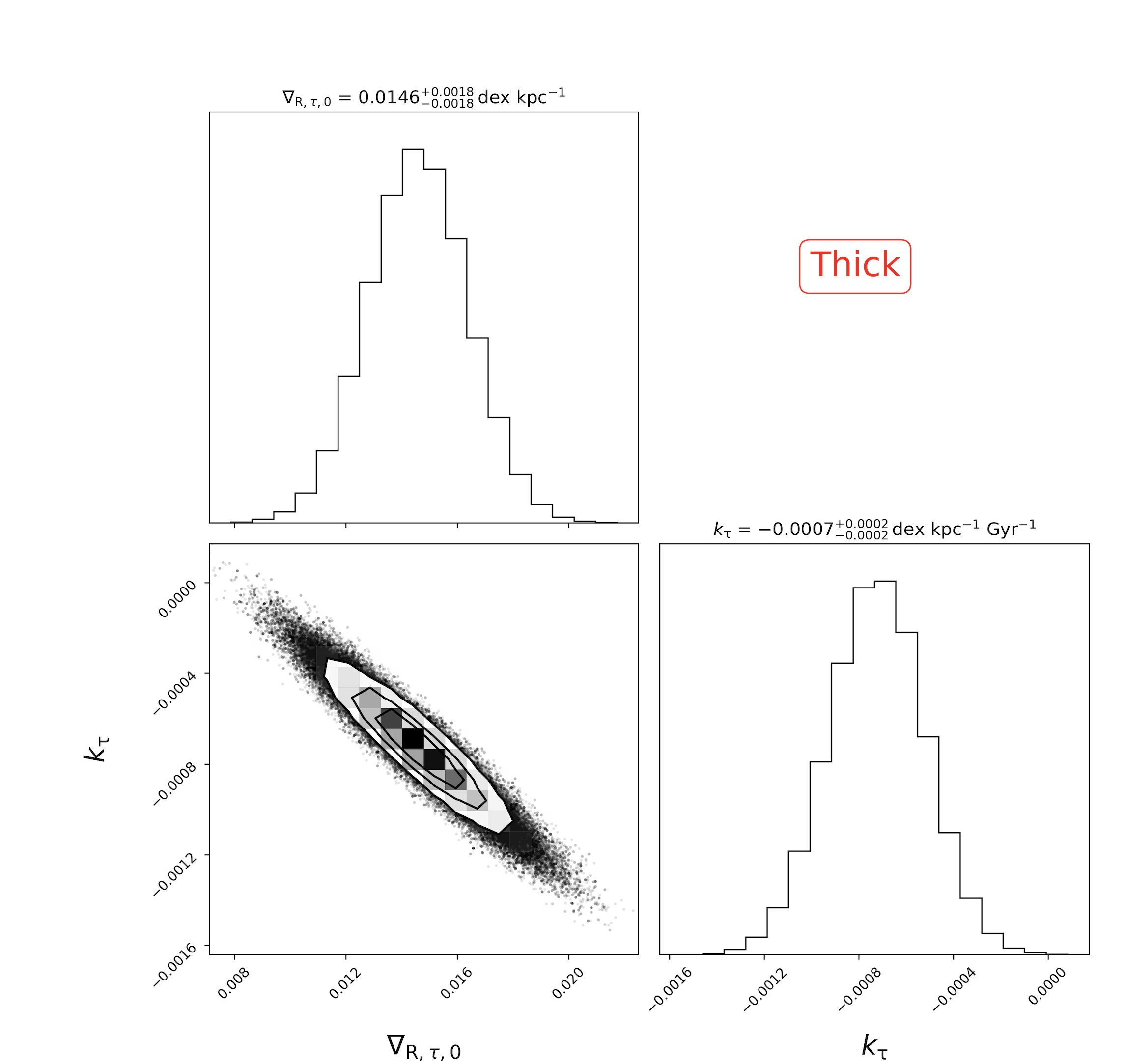}
}

\caption{Similar to Fig. C1, but for the age-dependent models of the thin (left panel) and thick (right panel) disks, parameterized as $\nabla_{\mathrm{R}}(\tau)$ = $\nabla_{\mathrm{R},\tau,0}$ $+$ $\Delta\nabla_{\mathrm{R},\tau}$\,$\tanh\left((\tau-\tau_{\rm c})/\sigma_{\tau}\right)$ and $\nabla_{\mathrm{R}} (\tau)$ = $\nabla_{R,\tau,0}$ $+$ $k_{\tau}$\,$\tau$, respectively.
The shadowed contours from inside to outside correspond to the confidence intervals of 1$\sigma$, 2$\sigma$ and 3$\sigma$, respectively.}
\end{figure*}
%%\label{Fig.C2}

\end{document}